\documentclass[aps,twocolumn,superscriptaddress]{revtex4}
\usepackage{graphicx,color,ulem}

\begin{document}

\title{First-order magnetic and structural phase transitions in Fe$_{1+y}$Se$_x$Te$_{1-x}$}

\author{Shiliang Li}
\affiliation{
Department of Physics and Astronomy, The University of Tennessee, Knoxville, Tennessee 37996-1200, USA
}
\author{Clarina de la Cruz}
\affiliation{
Department of Physics and Astronomy, The University of Tennessee, Knoxville, Tennessee 37996-1200, USA
}
\affiliation{
Neutron Scattering Science Division, Oak Ridge National Laboratory, Oak Ridge, Tennessee 37831-6393, USA
}
\author{Q. Huang}
\affiliation{
NIST Center for Neutron Research, National Institute of Standards and Technology, Gaithersburg, MD 20899, USA
}
\author{Y. Chen}
\affiliation{
NIST Center for Neutron Research, National Institute of Standards and Technology, Gaithersburg, MD 20899, USA
}
\author{J. W. Lynn}
\affiliation{
NIST Center for Neutron Research, National Institute of Standards and Technology, Gaithersburg, MD 20899, USA
}
\author{Jiangping Hu}
\affiliation{
Department of Physics, Purdue University, West Lafayette, IN 47907, USA}
\author{Yi-Lin Huang}
\affiliation{
Institute of Physics, Academia Sinica, Nankang, Taipei, Taiwan
}
\author{Fong-Chi Hsu}
\affiliation{
Institute of Physics, Academia Sinica, Nankang, Taipei, Taiwan
}
\author{Kuo-Wei Yeh}
\affiliation{
Institute of Physics, Academia Sinica, Nankang, Taipei, Taiwan
}
\author{Maw-Kuen Wu}
\affiliation{
Institute of Physics, Academia Sinica, Nankang, Taipei, Taiwan
}
\author{Pengcheng Dai}
\affiliation{
Department of Physics and Astronomy, The University of Tennessee, Knoxville, Tennessee 37996-1200, USA
}
\affiliation{
Neutron Scattering Science Division, Oak Ridge National Laboratory, Oak Ridge, Tennessee 37831-6393, USA
}

\begin{abstract}
We use bulk magnetic susceptibility, electronic specific heat, and neutron
scattering to study structural and magnetic phase transitions in Fe$_{1+y}$Se%
$_x$Te$_{1-x}$. Fe$_{1.068}$Te exhibits a first order phase transition near
67 K with a tetragonal to monoclinic structural transition and
simultaneously develops a collinear antiferromagnetic (AF) order responsible
for the entropy change across the transition. Systematic studies of FeSe$%
_{1-x}$Te$_x$ system reveal that the AF structure and lattice distortion in
these materials are different from those of FeAs-based pnictides. These
results call into question the conclusions of present density functional
calculations, where FeSe$_{1-x}$Te$_x$ and FeAs-based pnictides are expected
to have similar Fermi surfaces and therefore the same spin-density-wave AF
order.
\end{abstract}



\maketitle

\section{INTRODUCTION}
Superconductivity was recently discovered in $\alpha$-phase
FeSe$_x$ system \cite{mawkuen2}, shortly after the discovery of
superconductivity in FeAs-based pnictides \cite{kamihara,xhchen,gfchen,ren,rotter,hhwen}. The $T_{c}$ of the Fe$_{1+y}$Se$_{x}$Te$
_{1-x}$ system can reach up to 14 K at ambient pressure \cite{mawkuen2,mawkuen1,margadonna,fangmh} and 27 K at a pressure of 1.48 GPa \cite{mizuguchi}. Contrary to the earlier prediction of a low-$T_{c}$
conventional superconductor \cite{atzeri}, density functional calculations
of the electronic structure, magnetism and electron-phonon coupling for the
superconducting phase of Fe$_{1+y}$Se$_{x}$Te$_{1-x}$ suggest that
superconductivity in this class of materials is unconventional and mediated
by spin fluctuations \cite{subedi}. Furthermore, the calculated Fermi
surface of Fe$_{1+y}$Se$_{x}$Te$_{1-x}$ is very similar to that of the iron
pnictides such as LaFeAsO and SrFe$_2$As$_2$. If the observed collinear antiferromagnetic (AF) order in the
parent compounds of the FeAs-based pnictides \cite{clarina,zhao1,ychen,kimber,zhao3,qhuang,zhao2,goldman} is due to the spin-density-wave (SDW)
instability of a nested Fermi surface \cite{mazin,dong,ma,yin}, one would
expect to find the same AF structure or SDW instability in the
nonsuperconducting Fe$_{1+y}$Se$_{x}$Te$_{1-x}$. For FeAs-based materials such as LaFeAsO, CeFeAsO, and
PrFeAsO, neutron scattering experiments have shown that
the system exhibits a tetragonal to orthorhombic lattice distortion followed
by a collinear AF order with moment direction along the orthorhombic long ($
a $) axis (Fig. 1c) \cite{clarina,zhao1,ychen,kimber,zhao3}. In the case of Fe$
_{1+y}$Se$_{x}$Te$_{1-x}$, although Fe$_{1+y}$Te also undergoes a structural
distortion along with the establishment of a long-range AF order near 67 K 
\cite{fruchart,weibao}, the low temperature structure is monoclinic and the
magnetic structure can be either commensurate \cite{fruchart} or
incommensurate \cite{weibao}, much different from the commensurate AF
structure of FeAs-based pnictides.

To understand this apparent discrepancy, we carried out systematic neutron
scattering studies of the Fe$_{1+y}$Te system. We find that excess Fe ions
in Fe$_{1+y}$Te sitting in the octahedral sites \cite{fruchart,gronvold,ward}
have magnetic moments \cite{fruchart}. Although stoichiometric FeTe is
difficult to synthesize \cite{gronvold}, the Fe spins in Fe$_{1.068}$Te form
a collinear AF structure with moments confined within the $a$-$b$ plane of
the monoclinic structure as shown in Fig. \ref{structure}(b). Consistent
with earlier measurements \cite{tsubokawa,westrum}, we find that the AF
phase transition is first-order with an entropy change of $\sim $3.2 J/(mol$
\cdot $K). Systematic studies of FeSe$_{0.287}$Te$_{0.743}$ and FeSe$
_{0.568} $Te$_{0.432}$ reveal that the differences in lattice distortions
between FeSe$_{1-x}$Te$_{x}$ and LaFeAsO can account for the differences in
their magnetic structures. These results are difficult to explain within the
previous density functional calculations, where FeTe, FeSe, and LaFeAsO are
expected to have similar Fermi surfaces and therefore similar
SDW-type AF order \cite{subedi}. However, more recent density
functional calculations suggest that the excess Fe in Fe$_{1+y}$Te is
strongly magnetic and is also an electron donor, which might account for the
observed differences in magnetic structures of Fe$_{1+y}$Te and other
FeAs-based pnictides \cite{lijunzhang}.

\section{EXPERIMENTAL RESULTS and DISCUSSIONS}

We prepared powder samples of Fe$_{1+y}$Se$_{x}$Te$_{1-x}$ with nominal
composition of $x=0,0.3,0.5$ using the method described elsewhere \cite
{mawkuen1}. Fe$_{1+y}$Te is non-superconducting while the other two samples
have $T_{c}$ of $\sim $ 14 K. Powder neutron diffraction data were taken on
the BT-1 powder diffractometer at the NIST Center for Neutron Research
(NCNR). The nuclear and magnetic structure transitions were further studied
on the BT-7 triple-axis spectrometer at the NCNR. We define the nuclear wave
vector $Q$ at ($q_{x}$,$q_{y}$,$q_{z}$) as ($H$,$K$,$L$)=($q_{x}a/2\pi $,$
q_{y}b/2\pi $,$q_{z}c/2\pi $) reciprocal lattice units (r.l.u.) in both the
tetragonal and monoclinic unit cells. We used commercial SQUID and PPMS
systems to measure the DC susceptibility and specific heat of the samples
used for neutron measurements.

\begin{figure}[tbp]
\includegraphics[scale=.45]{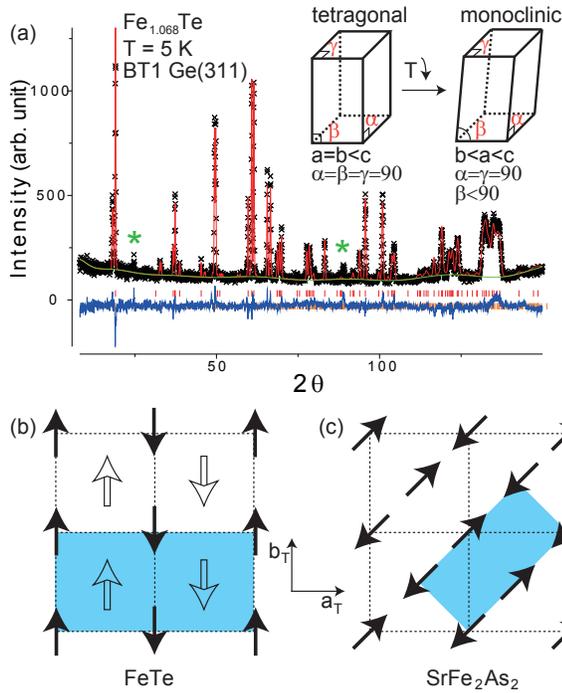}
\caption{ (a) Neutron powder diffraction data of Fe$_{1.068}$Te at $T = 5$ K
collected on the BT-1 diffractometer with Ge(3,1,1) monochromator and an
incident beam wavelength $\lambda =2.0785$ \AA . The lattice
structure is described by the monoclinic space group $P2_{1}/m$, which
changes to tetragonal $P4/nmm$ above $T_{N}$, as illustrated schematically
in the inset. (b) Schematic in-plane spin structure of Fe$_{1.068}$Te. The
solid arrows and hollow arrows represent two sublattices of spins, which can
be either parallel or antiparallel. The shaded area indicates the magnetic
unit cell. (c) Schematic in-plane spin structure of SrFe$_{2}$As$_{2}$ from
Ref. \cite{zhao2}. }
\label{structure}
\end{figure}

We first discuss the nuclear and magnetic structures of the
non-superconducting Fe$_{1+y}$Te. At 80 K, Fe$_{1+y}$Te has a tetragonal
crystal structure with space group $P4/nmm$ and no static magnetic order.
Our Rietveld analysis reveals that the system actually has excess Fe with $
y=0.068$. On cooling to 5 K, the nuclear structure changes to monoclinic
with the space group $P2_{1}/m$, as shown by the neutron powder diffraction
data in Fig. \ref{structure}(a) and refinement results in Table \ref{refinement}. For oxypinctides such as LaFeAsO and CeFeAsO, the lattice
distortion changes the symmetry from tetragonal to orthorhombic \cite
{clarina,zhao1}. In the case of Fe$_{1+y}$Te, the lattice distortion is from
tetragonal to monoclinic with the $\beta $ angle between $a$ and $c$ axis
being reduced to less than 90 degrees while the nearest Fe-Fe distance is
unchanged.

Figure \ref{structure}(c) shows the in-plane spin structure of LaFeAsO and SrFe$_{2}$As$
_{2}$, where the Fe moments form a collinear AF structure with spin
directions along the $a$-axis of the orthorhombic direction. This magnetic
structure appears to be ubiquitous for parent compounds of FeAs-based
superconductors \cite{clarina,zhao1,ychen,kimber,zhao3,qhuang,zhao2,goldman}. Our refinement on Fe$
_{1.068}$Te suggests that the spin structure in this system is also
collinear, and consists of two sublattices with a major component of the
moment along the tetragonal $b$ axis as shown in Fig. \ref{structure}(b).
Based on our powder diffraction data, we cannot conclusively determine the
relative spin directions between the two sublattices (they may be parallel
or antiparallel). The in-plane spin directions in Fe$_{1.068}$Te are rotated
45 degrees from those in the Fe-As materials. This is different from the
prediction of the density functional calculations \cite{subedi}, where Fermi
surfaces of these two materials are expected to be very similar.

In addition to the large moment (1.97 $\mu_B$) along the tetragonal $b$-axis
direction, we find that the projections of the moment along the $a$ and $c$
-axes were -0.56 $\mu_B$, and 0.25 $\mu_B$, respectively. Although the total
moment is similar in Fe$_{1.068}$Te and Fe$_{1.125}$Te, the $c$-axis
component of the moment in Fe$_{1.125}$Te is 1.36 $\mu_B$ \cite{fruchart}.
This difference may be related to the excess Fe ions in the octahedral sites 
\cite{fruchart,gronvold,ward}, which is expected to be strongly magnetic 
\cite{lijunzhang}. Since the moments of the excess Fe ions are randomly
distributed between the Fe-Fe layers, the Fe moments in Fe$_{1.125}$Te tend
to cant toward the $c$-axis. With further addition of excess Fe into the
system, the static antiferromagnetic order actually becomes incommensurate 
\cite{weibao}. The reduction of the excess Fe drives the system toward the
stoichiometric FeTe and decreases the influence of the excess Fe, this in
turn favors the commensurate AF spin structure in Fig. 1(b).

\begin{figure}[tbp]
\includegraphics[scale=.45]{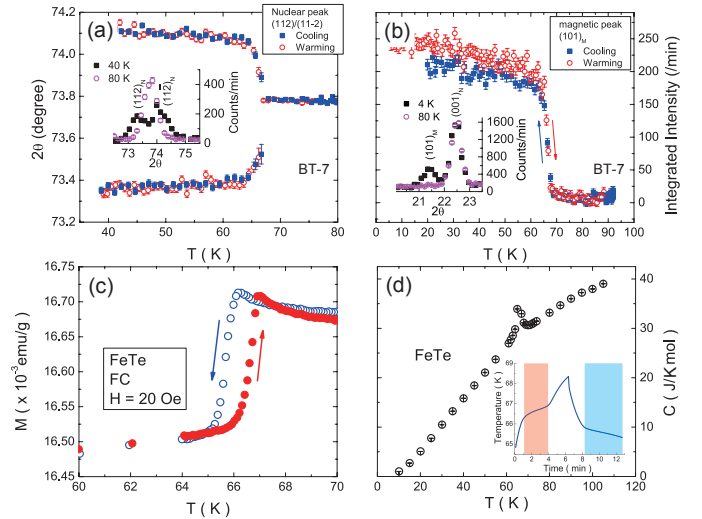}
\caption{(a) Splitting of the (1,1,2) and (1,1,-2) nuclear peaks with
decreasing temperature due to the tetragonal-monoclinic lattice distortion.
(b) Temperature dependence of the (101)$_{M}$ AF Bragg peak indicates a
strong coupling to the structural distortion. (c) DC magnetic susceptibility
measured by SQUID shows a temperature hysteresis of about 1 K. (d) Specific
heat measurements show a sharp peak around the structural/magnetic phase
transition. The inset shows the raw data of the thermal-relaxation
calorimeter of the PPMS, where the plateaus during warming and cooling
processes clearly reveals the absorption and liberation of the latent heat. }
\label{phasetransition}
\end{figure}

To understand the nuclear and magnetic phase transitions, we focus on the
(1,1,2)/(1,1,-2) nuclear and (1,0,1)$_{M}$ magnetic Bragg peaks. As shown in
the inset of Fig. \ref{phasetransition}(a), the (1,1,2)/(1,1,-2) reflections
split into two peaks due to the tetragonal-monoclinic structural transition.
By fitting with one and two Gaussian peaks at high and low temperatures,
respectively, we find that the structural phase transition happens near 67
K. Figure \ref{phasetransition}(b) shows that the temperature dependence of
the (1,0,1)$_{M}$ magnetic peak is clearly associated with the structural
phase transition. In addition, the full width at half maximum (FWHM) of the
(1,0,1)$_{M}$ peak is larger than the resolution due to the splitting of the
(1,0,1)$_{M}$ and (1,0,-1)$_{M}$ peaks.

To see if the 67 K phase transition is first or second order, we measured
the magnetic susceptibility using a SQUID. Figure \ref{phasetransition}(c)
shows that DC susceptibility with field-cooled (FC) process and an applied
magnetic field of 20 Oe has a clear hysteresis near the structural/magnetic
phase transition. The first-order nature of the structural/magnetic phase
transitions is shown unambiguously in the heat capacity measurement. Similar
to the previous study in Fe$_{1.11}$Te \cite{westrum}, a sharp peak is found
around the phase transition temperature [Fig. \ref{phasetransition}(d)].
Since the heat capacity option of the PPMS does not to work accurately in
the vicinity of the first-order transition \cite{lashley}, the inset of Fig. 
\ref{phasetransition}(d) shows the raw data of the calorimeter, where the
plateaus in the heating and cooling processes clearly reveal the heat
absorption and liberation during the first-order phase transition. Based on
these data, we estimate that the latent heat of the phase transition is $
\sim $215 J/mol, meaning a change of entropy of $\Delta S\sim 3.2$ J/(mol$
\cdot $K) through the transition. If we assume that the Fe moment in FeTe is
about 3.87 $\mu _{B}$ in the paramagnetic state \cite{komarek} and 1.7 $\mu
_{B}$ below $T_{N}$, the change of the entropy across the transition based
on an Ising model is about 3.5 J/(mol$\cdot $K). This result
suggests that the major contribution to the entropy change at the
phase transition arises from the spin ordering, which supports the
view that the first order phase transition is driven by the
magnetism \cite{cfang}.

\begin{figure}[tbp]
\includegraphics[scale=0.55]{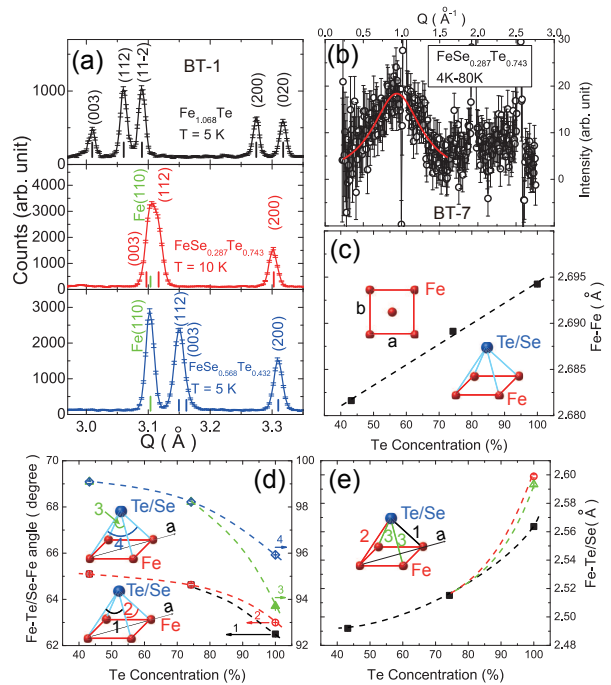}
\caption{ (a) Evolution of the (1,1,2)/(1,1,-2) and (2,0,0)/(0,2,0) peaks at
low temperatures in Fe$_{1.068}$Te, FeSe$_{0.287}$Te$_{0.743}$ and FeSe$
_{0.568}$Te$_{0.432}$. (b) Short-range AF fluctuations at $Q$ = 0.938 \AA $
^{-1}$ with FWHM = 0.67 \AA $^{-1}$ in FeSe$_{0.287}$Te$_{0.743}$. (c)-(e)
doping dependence of nearest-neighbor Fe-Fe distance, the angles of
Fe-Te/Se-Fe, and Fe-Te/Se distances, respectively. }
\label{doping}
\end{figure}

Finally, we discuss the lattice distortions and magnetic structure in Fe$
_{1+y}$Se$_{1-x}$Te$_{x}$ as superconductivity is induced by replacing Te
with Se \cite{mawkuen1,fangmh}. Although we find no static long range ordered
magnetic Bragg peaks in the superconducting FeSe$_{0.287}$Te$_{0.743}$ and
FeSe$_{0.568}$Te$_{0.432}$ similar to the Fe-As based materials \cite{clarina,zhao1,ychen,kimber,zhao3,qhuang,zhao2,goldman}, short-range spin fluctuations with correlation length of
9.4 \AA\ were found in FeSe$_{0.287}$Te$_{0.743}$ at $Q$ = 0.938 \AA $^{-1}$
as shown in Fig. \ref{doping}(b). The $Q$ value is slightly less than the $Q$
value 0.974 \AA $^{-1}$ at the commensurate position (1,0,1)$_{M}$, which is
likely due to the variation of the magnetic form factor as well as a
possible variation of the magnetic structure factor. However, one has to
be vigilant for impurity phases \cite{terzieff,komarek}. For example, we can
clearly see the strong (1,1,0) cubic Fe impurity peaks in Fig. \ref{doping}
(a) in both superconducting samples, which suggests the nonstoichiometry of
our samples and the possible existence of other phases. We therefore
cannot conclude unambiguously that these magnetic fluctuations originate in
the superconducting phase, although that is likely the case.

Figure \ref{doping} and Table I summarize the doping evolution of some
structural parameters for Fe$_{1+y}$Se$_{1-x}$Te$_x$. The nearest Fe-Fe
distance linearly increases with increasing Te concentration, as shown in
Fig. \ref{doping}(c). On the other hand, the Fe-Te/Se-Fe angles [Fig. \ref
{doping}(d)] decrease with increasing Te concentration. For Fe$_{1.068}$Te,
the Fe-Te/Se-Fe angle along the $b$ axis is much smaller than that along the 
$a$ axis. Because of the low-temperature monoclinic structure, the perfect
Fe-Te/Se tetrahedron is distorted, resulting in a slight difference in the
Fe-Te/Se-Fe angles between the nearest Fe ions, labeled as 1 and 2 in Fig. 
\ref{doping}(d). This distortion of the Fe-Te/Se tetrahedron is also
illustrated by the Fe-Te/Se distances, which increase with increasing Te
concentration [Fig. \ref{doping}(e)].

\begin{table}
\caption{\label{refinement}Refinement of powder diffraction data}
\begin{ruledtabular}
\begin{tabular}{lccccc}
\multicolumn{6}{l}{Fe$_{1.068}$Te(5 K),$P2_1/m$,$\chi^2$=1.559,$\beta=89.212(3)^\circ$}\\
\multicolumn{6}{l}{$a=3.83435(8)$(\AA),$b=3.78407(7)$(\AA),$c=6.25705(8)$(\AA)}\\
Atom & site & x & y & z & occupancy \\
Fe(1) & 2b & 0.75 & 0.25 & 0.0035(7) & 0.995(11) \\
Te & 2a & 0.25 & 0.25 & 0.2798(6) & 1\\
Fe(2) & 2a & 0.25 & 0.25 & 0.6812(5) & 0.068(3) \\
\hline
\multicolumn{6}{l}{Fe$_{1.068}$Te(80 K),$P4/nmm$,$\chi^2$=1.399}\\
\multicolumn{6}{l}{$a=3.81234(8)$(\AA),$b=3.81234(8)$(\AA),$c=6.2517(2)$(\AA)}\\
Atom & site & x & y & z & occupancy \\
Fe(1) & 2b & 0.75 & 0.25 & 0 & 0.995(11) \\
Te & 2a & 0.25 & 0.25 & 0.2829(4) & 1\\
Fe(2) & 2a & 0.25 & 0.25 & 0.7350(3) & 0.068(3) \\
\hline
\multicolumn{6}{l}{FeSe$_{0.287}$Te$_{0.743}$(10 K),$P4/nmm$,$\chi^2$=3.649}\\
\multicolumn{6}{l}{$a=3.8030(4)$(\AA),$b=3.8030(4)$(\AA),$c=6.0836(3)$(\AA)}\\
Atom & site & x & y & z & occupancy \\
Fe & 2b & 0.75 & 0.25 & 0 & 1 \\
Se & 2a & 0.25 & 0.25 & 0.2708(4) & 0.287\\
Te & 2a & 0.25 & 0.25 & 0.2708(4) & 0.743 \\
\hline
\multicolumn{6}{l}{FeSe$_{0.568}$Te$_{0.432}$(5 K),$P4/nmm$,$\chi^2$=1.985}\\
\multicolumn{6}{l}{$a=3.7924(1)$(\AA),$b=3.7924(1)$(\AA),$c=5.9551(3)$(\AA)}\\
Atom & site & x & y & z & occupancy \\
Fe & 2b & 0.75 & 0.25 & 0 & 1 \\
Se & 2a & 0.25 & 0.25 & 0.2715(3) & 0.568\\
Te & 2a & 0.25 & 0.25 & 0.2715(3) & 0.432\\
\end{tabular}
\end{ruledtabular}
\end{table}

To put these results in a proper context, we note that the density
functional calculations predicted a similar Fermi surface for Fe$_{1+y}$Te
and FeAs-based materials \cite{subedi}. Therefore, within the itinerant
electron picture where the observed AF order in these two classes of
materials arises from the same Fermi surface nesting, one should expect a
similar spin density wave instablilty, in contrast to the experimental
observation. However, the presence of a large magnetic moment on the excess
Fe ion in between the ordered Fe-layers in Fe$_{1+y}$Te might resolve this
inconsistency \cite{lijunzhang}. Alternatively, a model based on the
localized magnetic exchange interactions can well explain the experimental
results \cite{xiangtao}. In fact, the difference between the Fe-Te/Se-Fe
angles along the $a$ and $b$-axes should result in different next-nearest
neighbor couplings $J_{2a}$ and $J_{2b}$ that are responsible for the
observed collinear AF structure. The difference between angles 1 and 2 due
to the monoclinic lattice distortion may give rise to different nearest
neighbor couplings $J_{1a}$ and $J_{1b}$. This could in turn stabilize the
proposed spin structure in Fig. \ref{structure}(b) where a spin is actually
frustrated by the four nearest spins in a perfect rectangle \cite{cfang2}.

\section{CONCLUSIONS} 

In conclusion, we have systematically studied the structural and magnetic
phase transitions in $\alpha$-phase Fe$_{1+y}$Se$_x$Te$_{1-x}$ system. In
the pure Fe$_{1+y}$Te, we find structural and magnetic phase transitions are
intimately connected and first order in nature. The spin structure in Fe$
_{1+y}$Te is different from all other FeAs-based materials. Our results show
the important role played by the excess Fe ions in determining the magnetic
structure of Fe$_{1+y}$Te and suggest that the magnetic ordering can provide
enough energy for driving the first order phase transition. Our systematic
neutron diffraction measurements suggest that magnetic structure evolution
might be consistent with a local moment picture in Fe$_{1+y}$Se$_x$Te$_{1-x}$.

\section{acknowledgments}

We thank David Singh for helpful discussions.
This work is supported by the US NSF DMR-0756568, PHY-0603759, by the
US DOE, BES, through DOE DE-FG02-05ER46202 and Division of Scientific User
Facilities.


\end{document}